\documentstyle[sprocl]{article}

\def\al{\alpha}
\def\be{\beta}
\def\ga{\gamma}
\def\de{\delta}
\def\ep{\epsilon}

\def\th{\theta}

\def\ka{\kappa}
\def\la{\lambda}

\def\si{\sigma}

\def\ph{\phi}

\def\ps{\psi}

\def\Ga{\Gamma}

\def\mn{{\mu\nu}}

\def\cl{{\cal L}}

\def\prt{\partial}
\def\vev#1{\langle {#1}\rangle}

\def\fr#1#2{{{#1} \over {#2}}}
\def\frac#1#2{{\textstyle{{#1}\over {#2}}}}
\def\half{{\textstyle{1\over 2}}}
\def\lsim{\mathrel{\rlap{\lower4pt\hbox{\hskip1pt$\sim$}}
    \raise1pt\hbox{$<$}}}
\def\gsim{\mathrel{\rlap{\lower4pt\hbox{\hskip1pt$\sim$}}
    \raise1pt\hbox{$>$}}}
\def\sqr#1#2{{\vcenter{\vbox{\hrule height.#2pt
         \hbox{\vrule width.#2pt height#1pt \kern#1pt
         \vrule width.#2pt}
         \hrule height.#2pt}}}}

\def\lrprt{\stackrel{\leftrightarrow}{\partial}}

\def\lrDmu{\stackrel{\leftrightarrow}{D_\mu}}
\def\lrDnu{\stackrel{\leftrightarrow}{D^\nu}}

\def\Re{\hbox{Re}\,}

\newcommand{\beq}{\begin{equation}}
\newcommand{\eeq}{\end{equation}}
\newcommand{\bea}{\begin{eqnarray}}
\newcommand{\eea}{\end{eqnarray}}
\newcommand{\rf}[1]{(\ref{#1})}

\begin{document}

\title{Extending the Standard Model to Include CPT- and Lorentz-Breaking
Terms}

\author{DON COLLADAY}

\address{Physics Department, The College of Wooster, \\ Wooster,
OH 44691, USA} 


\maketitle\abstracts{ Low-energy remnant fundamental symmetry violations may
be present in nature at levels attainable in upcoming experiments.
These effects may arise through spontaneous symmetry breaking 
in a more complete Lorentz covariant theory 
underlying the standard model.
In this work the issue of parameterizing such violations in a 
consistent and complete manner is addressed.
The approach is to use the mechanism of
spontaneous symmetry breaking to generate all possible terms consistent with
gauge invariance and power-counting 
renormalizability to construct an extension of the standard model
that includes Lorentz- and CPT-breaking terms.
A consistent quantization of the theory is developed,
conventional quantum field theoretic techniques are shown to apply,
and some ramifications for quantum electrodynamics are explored.}

\section{Introduction and Motivation}

Lorentz invariance serves as a fundamental guiding
principle of virtually every theory describing fundamental particle 
interactions.  
When applied to local, point-particle theories, the assumed 
Lorentz invariance coupled with some mild technical assumptions 
leads one to conclude that CPT must be preserved in
the theory.\cite{cptthm}

However, if the fundamental theory underlying the standard model 
is constructed using nonlocal objects such as strings, 
Lorentz symmetry may not be exactly preserved in the low-energy 
limit of the full theory which is assumed to contain the standard model.
An explicit mechanism of this type 
has been proposed whereby spontaneous 
Lorentz symmetry breaking may occur in string theory.\cite{ks,kp1}

Our approach here is to use the mechanism of spontaneous symmetry breaking 
to generate a list of possible Lorentz violating interactions between 
standard model fields.
The standard model extension is constructed by selecting 
those terms satisfying SU(3)$\times$SU(2)$\times$U(1) gauge invariance
and power-counting renormalizability.\cite{ck1}
By only using the property of spontaneous symmetry breaking and not 
referring to explicit details of the underlying theory, we are able 
to construct a very general model of Lorentz breaking in the context
of the standard model.

High precision measurements involving atomic systems, 
\cite{ptrap,bkr}
clock comparisons, \cite{cc} and neutral meson oscillations 
\cite{kexpt}
provide stringent 
tests of Lorentz and CPT symmetry.
They are treated in great detail elsewhere in these proceedings.
The implications of CPT-violating terms of the type described in this 
work on baryogenesis have also been investigated.\cite{bert}

\section{Lorentz Violation via Spontaneous Symmetry Breaking}

Conventional spontaneous symmetry breaking occurs in the 
Higgs sector of the standard model 
where the Higgs field gains an expectation value, 
thereby partially 
breaking SU(2)$\times$U(1) gauge invariance.
This happens because an assumed potential for the Higgs field has a 
minimum at some nonzero value of the field.

For example, given a simple Lagrangian describing a single
fermion field $\psi$ and a single scalar field $\phi$ of the form
\beq
{\cal{L}} = {\cal{L}}_{0} - {\cal{L}}^{\prime} 
\quad ,
\eeq
where
\beq
{\cal L}^{\prime} \supset {\la} \phi
\overline{\psi} \psi + \rm{h.c.}
- (\phi^{2} - a^{2})^{2}
\quad ,
\label{higgsmech}
\eeq
a nonzero vacuum expectation value $\vev{\phi}$ for the 
scalar field will minimize the energy, 
hence generating a fermion mass of $m_{f}=\la \vev{\phi}$.
This expectation value of the scalar field breaks the 
SU(2)$\times$U(1) gauge invariance because $\vev{\phi}$ no
longer transforms nontrivially under the gauge group.
Lorentz symmetry is maintained in this case because $\vev{\phi}$
and $\phi$ both transform trivially under boosts and rotations.

Notice that if a \it tensor \rm $T$ gains a nonzero vacuum expectation value,
$\vev{T}$, Lorentz symmetry will be spontaneously broken.
To see how this might occur, consider a Lagrangian
describing a fermion $\psi$ and a tensor $T$ of the form
\beq
{\cal{L}} = {\cal{L}}_{0} - {\cal{L}}^{\prime} 
\quad ,
\eeq
where
\beq
{\cal L}^{\prime} \supset \fr {\la}{M^k} {T} \cdot \overline{\psi} 
\Ga (i \partial )^k \psi + \rm{h.c.} + V(T)
\quad .
\eeq
In this expression, $\la$ is a dimensionless coupling, 
$M$ is some heavy mass scale of the underlying theory,
$\Ga$ denotes a general gamma matrix structure in the Dirac algebra,
and $V(T)$ is a potential for the tensor field. (indices are
suppressed for notational simplicity)
The potential $V(T)$ is assumed to arise from a theory underlying
the standard model.  
Terms contributing to $V(T)$ are precluded from
conventional renormalizable four-dimensional field theories,
but may arise in the low-energy limit 
of a more general theory such as string theory.
\cite{ks}

If the potential $V(T)$ is such that it has a nontrivial minimum, a 
vacuum expectation value $\vev{T}$ will be generated.
There will then be a term of the form
\beq
{\cal L}^{\prime} \supset \fr {\la}{M^k} \vev{T} \cdot \overline{\psi} 
\Ga (i \partial )^k \chi + \rm{h.c.} 
\quad ,
\label{vevt}
\eeq
present in the Lagrangian.
Terms of this type can break Lorentz invariance and various 
discrete symmetries C, P, T, CP, and CPT.

\section{Relativistic Quantum Mechanics and Field Theory}
To develop theoretical techniques for treating generic terms of the 
form given in Eq.~\rf{vevt}, a specific example is studied.
The example presented here involves a single fermion Lagrangian
containing Lorentz-violating terms without derivative couplings 
($k=0$) that also violate CPT.

We proceed by listing the possible gamma-matrix structures that could 
arise in such a term:
\beq
\Ga \sim \{1, \ga^{\mu}, \ga^5 \ga^{\mu}, 
\sigma^{\mn}, \ga^5\}
\quad .
\label{gambasis}
\eeq
The condition that a fermion bilinear with no derivative couplings 
violates CPT is equivalent to the requirement that $\{\Ga,\ga^{5}\}=0$.
Half of the matrices in Eq. \rf{gambasis} 
satisfy this condition: $\Ga \sim \ga^{\mu}$ and 
$\Ga \sim \ga^{5}\ga^{\mu}$.
The contribution to the lagrangian from these terms can be written as
\beq
{\cal L}_{a}^{\prime} \equiv a_{\mu} \overline{\psi} \ga^{\mu} \psi \quad , 
\quad
{\cal L}_{b}^{\prime} \equiv b_{\mu} \overline{\psi} \ga_5 \ga^{\mu} \psi 
\quad ,
\label{abmu}
\eeq
where $a_{\mu}$ and $b_{\mu}$ are constant coupling coefficients that 
parameterize the tensor expectation values and relevant couplings 
arising in Eq.~\rf{vevt}.
These parameters are assumed suppressed with respect to other 
physically relevant energy scales in the low-energy effective theory.

Combining these terms with the conventional single fermion terms
gives a model lagrangian of
\beq
{\cal L} = \fr i 2 \overline{\ps} \ga^{\mu} \lrprt_\mu \psi 
- a_{\mu} \overline{\psi} \ga^{\mu} \psi 
- b_{\mu} \overline{\psi} \ga_5 \ga^{\mu} \psi 
- m \overline{\psi} \psi
\quad .
\label{modlag}
\eeq
Several features of this modified theory are immediately apparent.
The lagrangian is hermitian, thereby obeying conventional quantum 
mechanics, conservation of probability and unitarity.
Translational invariance implies the existence of a conserved energy 
and momentum.
Explicitly, the conserved four-momentum is constructed as
\beq
P_\mu = \int d^3 x \Theta^0_{\ \mu} = 
\int d^3 x \frac 1 2 i \overline{\psi} \ga^0 \lrprt_\mu \psi
\quad ,
\label{emom}
\eeq
just as in the conventional case.
The Dirac equation that results from Eq.~\rf{modlag} is linear in the 
fermion field allowing an exact solution of the free theory.
Finally, the global U(1) invariance of the model lagrangian implies 
the existence of a conserved current 
$j_{\mu}=\overline{\psi}\ga^{\mu} \psi$.

The resulting Dirac equation obtained by variation of Eq.~\rf{modlag} 
with respect to the fermion field is
\beq
(i \ga^{\mu} \partial_{\mu} - a_\mu \ga^\mu - 
b_{\mu} \ga_5 \ga^\mu - m) \psi = 0
\quad .
\eeq
Due to the linearity of the equation, plane-wave solutions 
\beq
\psi(x) = e^{\pm i p_{\mu} x^{\mu}} w(\vec{p}) 
\quad ,
\eeq
can be used to solve the equation.
Substitution into the modified Dirac equation yields
\begin{eqnarray}
(\pm p_{\mu} \ga^{\mu} - a_{\mu} \ga^{\mu} - 
b_{\mu} \ga_5 \ga^{\mu} - m) w(\vec{p}) &\equiv& M_{\pm} w(\vec{p}) 
\nonumber \\ &=& 0 
\quad .
\end{eqnarray}
A nontrivial solution will exist only if $Det M_{\pm} = 0$.
This imposes a condition on $p^0(\vec{p}) \equiv E(\vec{p})$
generating a dispersion relation for the fermion.

The general solution involves a fourth-order polynomial 
that can be solved algorithmically, but the resulting
solution is complex and not very illuminating.
Here we will consider the special case $\vec{b}=0$.
In this case the exact dispersion relations are
\beq
E_{+}(\vec{p}) = \left[ m^2 + (|\vec{p} - \vec{a}| \pm b_0)^2 \right]^{1/2} 
+ a_0 \quad ,
\eeq
\beq
E_{-}(\vec{p}) = \left[ m^2 + (|\vec{p} + \vec{a}| \mp b_0)^2 \right]^{1/2} 
- a_0 \quad .
\eeq
Examination of the above energies reveals some qualitative effects of the 
CPT-violating terms.
The usual four-fold energy degeneracy of spin-$\frac 1 2$ 
particles and antiparticles 
is removed by $a_{\mu}$ and $b_{0}$.
The particle-antiparticle degeneracy is broken by $a_{\mu}$ and the 
helicity degeneracy is split by $b_{0}$.
The corresponding spinor solutions $w(\vec p)$ can be explicitly calculated, 
forming an orthogonal basis of states as expected.

One interesting feature of these solutions is the unconventional 
relationship that exists between momentum and velocity.
For a wave packet of positive helicity particles with four momentum 
$p^{\mu}=(E,\vec{p})$, the expectation value of the velocity operator 
$\vec{v} = i[H,\vec{x}] = \ga^0 \vec{\ga}$ is calculated as
\beq
\vev{\vec{v}} = 
\vev{ \fr {(|\vec{p} - \vec{a}| - b^0)} {(E - a^0)}
         \fr {(\vec{p} - \vec{a})} {|\vec{p} - \vec{a}|} }
         \quad .
\eeq
Examination of the velocity using a general dispersion relation
reveals that $|v_{j}| < 1$ for arbitrary $b_{\mu}$,
and that the limiting velocity as $\vec{p} \rightarrow \infty$ is 1.
This implies that the effects of the CPT violating terms are mild 
enough to preserve causality.
This will be verified independently from the perspective of field 
theory that we will now develop.

To quantize the theory, the general expansion for $\psi$ in terms of 
its spinor components given by
\begin{eqnarray}
\ps (x) & = & \int \fr {d^3 p} {(2 \pi )^3} \sum_{\al = 1}^{2} \left[
\fr m {E_u^{(\al)}} b_{(\al)} (\vec{p})
e^{-i p_u^{(\al)} \cdot x} u^{(\al)} (\vec{p}) \right. \nonumber \\ 
&& \left. \qquad \qquad \qquad \qquad
+ \fr m {E_v^{(\al)}} d^*_{(\al)} (\vec{p}) 
e^{i p_v^{(\al)} \cdot x} v^{(\al)} (\vec{p}) \right] 
\quad ,
\end{eqnarray}
is promoted in the usual way to an operator acting on a Hilbert space 
of basis states.

Calculation of the energy from Eq.~\rf{emom} using conventional 
normal ordering
yields a positive definite quantity (for $|a^{0}| <m$)
provided the following nonvanishing anticommutation relations are imposed
on the creation and annihilation operators:
\begin{eqnarray}
\{b_{(\al)} (\vec{p}), b^{\dagger}_{(\al^{\prime})} (\vec{p}^{~\prime}) \} 
& = & (2 \pi)^3
\fr {E_u^{(\al)}} {m}
\de_{\al \al^{\prime}}
\de^3 (\vec{p} - \vec{p}^{~\prime})
\quad ,
\nonumber \\
\{d_{(\al)} (\vec{p}), d^{\dagger}_{(\al^{\prime})} (\vec{p}^{~\prime}) \} 
& = & (2 \pi)^3
\fr {E_v^{(\al)}} {m}
\de_{\al \al^{\prime}}
\de^3 (\vec{p} - \vec{p}^{~\prime})
\quad .
\end{eqnarray}
The resulting equal-time anticommutators of the fields are
\begin{eqnarray}
\{ \psi_{\al}(t,\vec{x}), \psi_{\be}^{\dagger}(t,\vec{x}^{\prime})\} 
& = & \de_{\al \be} \de^{3} (\vec{x} - \vec{x}^{\prime})
\quad , 
\nonumber \\ 
\{ \psi_{\al}(t,\vec{x}), \psi_{\be}(t,\vec{x}^{\prime})\} & = & 0 
\quad ,
\nonumber \\
\{ \psi_{\al}^{\dagger}(t,\vec{x}), \psi_{\be}^{\dagger}(t,\vec{x}^{\prime})\} 
& = & 0
\quad .
\end{eqnarray}
These relations show that the conventional Fermi statistics remain 
unaltered.

The conserved charge $Q$ and momentum $P^{\mu}$ are computed as
\begin{eqnarray}
Q & = & \int \fr {d^3 p} {(2 \pi)^3}
\sum_{\al = 1}^2 \left[
\fr m {E_u^{(\al)}}
b^{\dagger}_{(\al)} (\vec{p}) b_{(\al)} (\vec{p}) - \fr m {E_v^{(\al)}}
d^{\dagger}_{(\al)} (\vec{p}) d_{(\al)} (\vec{p}) \right] \quad ,\\
P_{\mu} & = & \int \fr {d^3 p} {(2 \pi)^3} \sum_{\al = 1}^2 \left[
\fr m {E_u^{(\al)}} p^{(\al)}_{u \mu}
b^{\dagger}_{(\al)} (\vec{p}) b_{(\al)} (\vec{p}) \right. 
\nonumber \\ 
& & \qquad \qquad \qquad \qquad \qquad \left.
+ \fr m {E_v^{(\al)}} p^{(\al)}_{v \mu}
d^{\dagger}_{(\al)} (\vec{p}) d_{(\al)} (\vec{p}) \right]
\quad .
\end{eqnarray}
From these expressions we see that the charge of the fermion is 
unperturbed and the energy and momentum satisfy the same energy 
momentum relations that we found using relativistic quantum mechanics.

Causality is governed by the unequal-time anticommutation relations 
for the fermion fields.  
Explicit integration for the case of $\vec{b}=0$ proves that
\beq
\{\psi_{\al}(x), \overline{\psi}_{\be}(x^{\prime})\} = 0
\quad ,
\eeq
for spacelike separations $(x - x^{\prime})^2 < 0$.
This result shows that physical observables separated by spacelike 
intervals will in fact commute (for case $\vec b = 0$).
This agrees with our previous results obtained by examination of the 
velocity using the relativistic quantum mechanics approach.

Next the problem of extending the free field theory to interacting 
theory is addressed.
Most of the conventional formalism developed for perturbative 
calculations in the interacting theory carries over directly to the 
present case.
The main reason that these techniques work is that the Lorentz 
violating modifications introduced are linear in the fermion fields.
The main result is that the usual Feynman rules apply provided that 
the Feynman propagator is modified as
\beq
S_F(p) = \fr i {p_{\mu} \ga^{\mu} - a_{\mu} \ga^{\mu} - 
b_{\mu} \ga_5 \ga^{\mu} - m}
\quad ,
\eeq
and the exact spinor solutions of the modified free fermion theory 
are used on external legs.

\section{Standard Model Extension}

We now turn to the question of how to apply spontaneous symmetry
breaking to generate Lorentz-violating terms using standard model
fields.
Our approach is to consider all possible terms that can arise from
spontaneous symmetry breaking that satisfy power-counting 
renormalizability and preserve the SU(3)$\times$SU(2)$\times$U(1) 
gauge invariance of the standard model.\cite{ck1}
The relevant terms contribute to all sectors of the standard 
model.
In listing the terms below, the Lorentz violating terms are 
classified according to their properties under the CPT transformation.

In the lepton sector the 
left- and right-handed multiplets are of the form
\beq
L_A = \left( \begin{array}{c} \nu_A \\ l_A \end{array} \right)_L
\quad , \quad
R_A = (l_A)_R
\quad ,
\eeq
where $A = 1, 2, 3$ labels the flavor:
\beq
l_A \equiv (e, \mu, \tau) \quad , \quad
\nu_A \equiv (\nu_e, \nu_{\mu}, \nu_{\tau}) 
\quad .
\eeq

The Lorentz-violating terms satisfying the required properties are
\begin{eqnarray}
\cl^{\rm CPT-even}_{\rm lepton} &=& 
\half i (c_L)_{\mu\nu AB} \overline{L}_A \ga^{\mu} \lrDnu L_B
\nonumber\\ &&
+ \half i (c_R)_{\mu\nu AB} \overline{R}_A \ga^{\mu} \lrDnu R_B
\quad , 
\\ && \nonumber \\
\cl^{\rm CPT-odd}_{\rm lepton} & = &
- ({a}_{L})_{\mu A B} ~ \overline{L}_A \ga^{\mu} L_B \nonumber \\ 
& & - ({a}_{R})_{\mu A B} ~ \overline{R}_A \ga^{\mu} R_B 
\quad .
\label{leptonlv}
\end{eqnarray}
In the above expression
$c_{\mu\nu}$ and $a_{\mu}$ are constant coupling coefficients related 
to the background expectation values of the corresponding tensor 
fields, and $D^{\mu}$ is the usual covariant derivative.

These are not the final form of the standard model terms because the 
SU(2)$\times$U(1) symmetry is broken by the Higgs mechanism.  
Once this breaking occurs, the fields in Eq.~\rf{leptonlv}
can be rewritten in terms of the physical Dirac spinors corresponding 
to the observed leptons and neutrinos.
For example, the CPT-odd lepton terms become
\begin{eqnarray}
\cl^{\rm CPT-odd}_{\rm lepton} & = &
-(a_\nu)_{\mu A B} ~ \overline{\nu}_{A}
\frac 1 2 (1 + \ga_5)\ga^{\mu} \nu_B \nonumber \\
& & - (a_l)_{\mu A B} ~ \overline{l}_A \ga^{\mu} l_B \nonumber \\ 
& & - (b_l)_{\mu A B} ~ \overline{l}_A \ga_5 \ga^{\mu} l_B 
\quad .
\label{lepcpt}
\end{eqnarray}
Note that $b_{\mu}$ coupling coefficients arise in 
the process of combining right- and left-handed fields into Dirac 
spinors.

If we now examine the first generation electron contribution 
corresponding to $A=B=1$, we find
\beq
\cl^{\rm CPT-odd}_{\rm lepton} 
\supset - (a_l)_{\mu 1 1}~\overline{e} \ga^{\mu} e 
- (b_l)_{\mu 1 1} ~\overline{e}\ga_5 \ga^{\mu} e 
\quad .
\eeq
These terms are exactly the form of Eq.~\rf{abmu}
that were analyzed in the previous section.
The relativistic quantum mechanics and field theoretic techniques 
developed to handle these terms are therefore directly applicable to 
electrons.
Terms in Eq. \rf{lepcpt} of the form $A \ne B$ lead to small lepton flavor-changing 
amplitudes.

The construction of the extension in the quark sector is 
similar to that in the lepton sector. 
The main difference is that corresponding right-handed quark fields 
are now present for each left-handed field unlike the case in 
the lepton sector.
The left- and right-handed quark multiplets are
\beq
Q_A = \left( \begin{array}{c} u_A \\ d_A \end{array} \right)_L
\quad , \quad
\begin{array}{c}
U_A = (u_A)_R \\
D_A = (d_A)_R
\end{array}
\quad ,
\eeq
where $A = 1, 2, 3$ labels quark flavor
\beq
u_A \equiv (u,c,t) \quad , \quad
d_A \equiv (d,s,b) \quad .
\eeq
The Lorentz-violating terms are of the same form as in the lepton 
sector and will not be explicitly given here.
The diagonal $A=B$ terms are again of the same form as Eq.~\rf{abmu}
analyzed in the previous section.
The quark $a_{\mu}$ terms are particularly interesting because they can lead 
to CPT-violating effects in neutral meson systems.\cite{kost}

Turning next to the Higgs sector, there are contributions involving two Higgs 
fields, and generalized Yukawa coupling terms involving a single Higgs 
and two fermion fields.
The Lorentz-violating terms quadratic in the Higgs fields are
\begin{eqnarray}
\cl^{\rm CPT-even}_{\rm Higgs} &=&
\half (k_{\ph\ph})^{\mu\nu} (D_\mu\ph)^\dagger D_\nu\ph 
+ {\rm h.c.}
\nonumber\\ &&
-\half (k_{\ph B})^{\mu\nu} \ph^\dagger \ph B_{\mu\nu}
\nonumber\\ &&
-\half (k_{\ph W})^{\mu\nu} \ph^\dagger W_{\mu\nu} \ph 
\quad ,
\end{eqnarray}
\beq
\cl^{\rm CPT-odd}_{\rm Higgs}
= i (k_\ph)^{\mu} \ph^{\dagger} D_{\mu} \ph + {\rm h.c.} 
\quad ,
\eeq
where $W_{\mu\nu}$ and $B_{\mu\nu}$ are the field strengths for the 
SU(2) and U(1) gauge fields and the various $k$ parameters are 
coupling constants related to tensor expectation values.

The Yukawa type terms involving one Higgs are
\begin{eqnarray}
\cl^{\rm CPT-even}_{\rm Yukawa} = 
&& - \half 
\left[
(H_L)_{\mu\nu AB} \overline{L}_A \ph \si^{\mu\nu} R_B
\right.
\nonumber\\ &&
\left.
+(H_U)_{\mu\nu AB} \overline{Q}_A \ph^c \si^{\mu\nu} U_B 
\right.
\nonumber\\ &&
\left.
+(H_D)_{\mu\nu AB} \overline{Q}_A \ph \si^{\mu\nu} D_B
\right]
+ {\rm h.c.}
\quad ,
\end{eqnarray}
where the $H$ parameters are related to tensor expectation values.

One interesting result of including these terms into the standard 
model is that the conventional SU(2)$\times$U(1) breaking is modified.
When the static potential is minimized, the $Z^{0}$ boson gains an 
expectation value of
\beq 
\vev{Z_\mu^0} = \fr 1 q {\sin2\th_W} 
(\Re \hat k_{\ph\ph})^{-1}_{\mu\nu} k^\nu_\ph
\quad ,
\eeq
where $\hat k_{\ph\ph}^{\mu\nu} = \eta^{\mu\nu} + 
k_{\ph\ph}^{\mu\nu}$, $q$ is the electric charge,
and $\th_{W}$ is the weak mixing angle.
Note that if the CPT-odd term $k_{\ph}$ vanishes then 
$\vev{Z_\mu^0}=0$.  This is reasonable because a nonzero value of 
$\vev{Z_\mu^0}$ violates CPT symmetry.

The final sector to be examined is the gauge sector.  
The various Lorentz-breaking terms satisfying our criteria are
\begin{eqnarray}
\cl^{\rm CPT-even}_{\rm gauge} &=&
-\half (k_G)_{\ka\la\mu\nu} {\rm Tr} (G^{\ka\la}G^{\mu\nu})
\nonumber\\ &&
-\half (k_W)_{\ka\la\mu\nu} {\rm Tr} (W^{\ka\la}W^{\mu\nu})
\nonumber\\ &&
-\frac 1 4 (k_B)_{\ka\la\mu\nu} B^{\ka\la}B^{\mu\nu}
\quad ,
\label{gaugee}
\end{eqnarray}
\begin{eqnarray}
\cl^{\rm CPT-odd}_{\rm gauge} & = &
k_{3\ka} \ep^{\ka\la\mu\nu}
{\rm Tr} (G_\la G_{\mu\nu} + \frac {2i} 3 G_\la G_\mu G_\nu) 
\nonumber \\ 
& +& k_{2\ka} \ep^{\ka\la\mu\nu}
{\rm Tr} (W_\la W_{\mu\nu} + \frac {2i} 3 W_\la W_\mu W_\nu) 
\nonumber \\ 
& +& k_{1\ka} \ep^{\ka\la\mu\nu} B_\la B_{\mu\nu} 
\quad .
\label{gaugeo}
\end{eqnarray}
In these expressions, the $k$ terms are the background coupling constants
and the $G^{\mu\nu}$, $W^{\mu\nu}$, and $B^{\mu\nu}$ are the field 
strengths for the SU(3), SU(2), and U(1) gauge fields respectively.

The CPT-odd terms can generate negative contributions to the energy 
\cite{cfj} creating an instability in the theory.  
One option is to set these coefficients to zero, and show that they 
remain zero at the quantum level.  This has been carried out to the 
one-loop level by utilizing an anomaly cancellation mechanism that 
must be inherited from any consistent theory underlying the standard 
model.\cite{ck1}

\section{Restriction to QED}

Here we restrict our attention to the theory of electrons and photons 
that results from the above extension of the standard model.
The usual QED Lagrangian is 
\beq
\cl^{\rm QED}_{\rm electron} = 
\half i \overline{\ps} \ga^\mu \lrDmu \ps 
- m_e \overline{\ps} \ps
- \frac 1 4 F_{\mu\nu}F^{\mu\nu}
\quad ,
\eeq
where $\psi$ is the electron field, $m_{e}$ is its mass,
and $F^{\mu\nu}$ is the photon field strength tensor.

Selecting out the CPT-even electron terms that violate Lorentz symmetry
from the full standard model extension yields
\begin{eqnarray}
\cl^{\rm CPT-even}_{\rm electron} &=& 
- \half H_{\mu\nu} \overline{\ps} \si^{\mu\nu} \ps 
\nonumber\\ &&
+ \half i c_{\mu\nu} \overline{\ps} \ga^{\mu} \lrDnu \ps 
\nonumber\\ &&
+ \half i d_{\mu\nu} \overline{\ps} \ga_5 \ga^\mu \lrDnu \ps
\quad ,
\end{eqnarray}
where $H$, $c$, and $d$ are constant coupling coefficients.

The CPT-odd electron terms are
\beq
\cl^{\rm CPT-odd}_{\rm electron} = 
- a_{\mu } \overline{\ps} \ga^{\mu} \ps 
- b_{\mu} \overline{\ps} \ga_5 \ga^{\mu}\ps 
\quad ,
\eeq
where $a$ and $b$ are parameters analogous to those in Eq.~\rf{abmu} 
applied to the electron field.

The photon corrections are given by
\beq
\cl^{\rm CPT-even}_{\rm photon} =
-\frac 1 4 (k_F)_{\ka\la\mu\nu} F^{\ka\la}F^{\mu\nu}
\quad ,
\eeq
and
\beq
\cl^{\rm CPT-odd}_{\rm photon} =
+ \half (k_{AF})^\ka \ep_{\ka\la\mu\nu} A^\la F^{\mu\nu}
\quad .
\eeq
The parameters $k_{F}$ and $k_{AF}$ are the appropriate linear 
combinations of parameters in Eqs.\rf{gaugee} and \rf{gaugeo} that 
result when the photon is defined as the unbroken U(1) electric force 
mediator.

For an explicit example, we will examine a special case in which
$(k_{AF})^\mu = 0$ (no CPT-odd piece), and 
$(k_F)_{0j0k} = - \half \be_j\be_k$.
With this choice the free photon lagrangian takes the form
\beq
\cl^{\rm special}_{\rm photon} 
= \half ( \vec{E}^2 - \vec{B}^2) 
+ \half (\vec\be \cdot \vec E)^2
\quad .
\eeq
Variation with respect to the dynamical fields yields 
the following modified source-free Maxwell equations:
\begin{eqnarray}
\vec{\nabla} \cdot \vec{E} 
& = & - \vec{\be} \cdot \vec{\nabla}
(\vec{\be} \cdot \vec{E})
\quad ,
\nonumber \\
\vec{\nabla} \times \vec{B} - \prt_0 \vec{E} 
& = & \vec{\be}~ \prt_0 (\vec{\be} \cdot \vec{E})
\quad .
\end{eqnarray}
The other two remain unmodified as they are simply a result of the 
definitions of $\vec{E}$ and $\vec{B}$ in terms of $A^{\mu}$ which 
have not been modified.

These equations can be solved with the plane-wave ansatz
\beq
A_\mu (x) \equiv 
\ep_\mu (p) \exp(-ip_\al x^\al) 
\quad ,
\eeq
because the modifications are linear in the dynamical fields.
Proceeding in the Lorentz gauge where $p_{\mu} A^{\mu} = 0$ (there is 
no difficulty in selecting this gauge because we have maintained 
gauge invariance in the standard-model extension) we find that a 
solution exists provided $p_{\mu}$ satisfies one of
\begin{eqnarray}
(p_o)^2 & = & 0 
\quad , \\
(p_e)^2 & = & - \fr { (\vec{\be}\times \vec p_e)^2}
{1 + \vec{\be}^2}
\quad ,
\end{eqnarray}
where $p_{o}$ denotes an ordinary mode 
and $p_{e}$ denotes an extraordinary mode of propagation.
The ordinary mode satisfies the conventional dispersion relation of 
electromagnetic waves and therefore behaves identically the same as 
ordinary photons.
The extraordinary mode is more interesting because it satisfies a 
modified dispersion relation.

Taking the case $\be \cdot \vec{p} = 0$ for simplicity, 
the ordinary mode is polarized with $\vec{A}_{o}$ along the direction 
of $\vec{p} \times \vec{\be}$ while the extraordinary mode 
$\vec{A}_{e}$ is polarized along $\vec{\be}$.
They are both perpendicular to the momentum of the wave
$\vec{p}$.
The group velocities defined by $\vec v_g \equiv \vec \nabla_p p^0$ 
take the form
\beq
\vec v_{g,o} = \hat p 
\quad , \quad
\vec v_{g,e} = 
\fr 1 {\sqrt{1 + \vec{\be}^2}} ~ \hat p 
\quad .
\eeq
The extraordinary mode travels with a modified velocity that is 
slightly less than the velocity of the ordinary mode.

A general wave will be a linear superposition of the $A_{o}$ and $A_{e}$ 
modes.
The electric field that results is
\beq
\vec{E}(t, \vec x) = 
-p^0 \left( c_o \hat{A}_o \sin[p^0 (r - t)]
+ c_e \hat{A}_e \sin [p^0 (\sqrt{1 + \vec{\be}^2}~ r - t)] \right) 
\quad ,
\eeq
where the weights $c_o$ and $c_{e}$ are fixed 
by the initial polarization conditions.
As the wave evolves in time, a
plane polarized wave will in general become elliptically polarized
after traveling a distance 
\beq
r \simeq \fr {\pi} 
{2 \left( \sqrt{1 + \vec{\be}^2} - 1 \right) p^0}
\simeq \fr {\pi} {\vec{\be}^2 p^{0}}
\quad ,
\eeq
where the approximation holds for $\vec{\be}^2 \propto k_{F} << 1$.
The magnetic field behaves similarly.
Terms of this form have interesting implications for photon birefringence,
in particular they contribute to polarization rotation from
quasars.\cite{pk}

\section{Summary}
A framework was 
presented that incorporates Lorentz- and CPT-violating effects into 
the context of conventional quantum field theory.
Using a generic spontaneous symmetry breaking mechanism as the source 
of these terms, an extension of the standard model that includes
Lorentz and CPT breaking was developed.
The extension preserves power-counting renormalizability and 
SU(3)$\times$SU(2)$\times$U(1) 
gauge invariance.
The parameters introduced can be used to establish quantitative bounds 
on CPT- and Lorentz-breaking effects in nature.
The restriction to QED was explored resulting in interesting implications 
for photon propagation.

\section*{Acknowledgments}
This work was supported in part by the United States Department of 
Energy under grant number DE-FG02-91ER40661.

\end{document}